# DNA Electromagnetic Properties and Interactions

M. H. S. Bukhari[1, 2#], S. Batool[3], Y. Raza[3], A. Bukhari[5], F. Memon[5], T. Razzaki[3], A. Rizvi[1], M. A. Rauf[4], and O. Bagasra[6]

[1] *Departments of Physics and Dentistry, Jazan University, Gizan 45142, Jazan, Saudi Arabia*
[2] *Malir University of Science & Technology, Malir, Karachi 75050, Pakistan.*

[3] *Departments of Molecular Pathology and the Stem Cells Research Laboratory, Sindh Institute of Urology and Transplantation, Karachi, Pakistan*

[4] *Department of Biotechnology, Aligarh Muslim University, Aligarh, India*

[5] *Liaquat University of Medical & Health Sciences, Jamshoro, Pakistan*

[6] *The South Carolina Center for Biotechnology, Claflin University, 400 Magnolia St, Orangeburg, SC 29115, USA*

# Corresponding author. Email: mbukhari@jazanu.edu.sa

## ABSTRACT

DNA is an essential molecule central to the survival and propagation of life, it was imperative to investigate possible electromagnetic properties inherent to it, such as existence of any non-trivial interactions of this molecule with electromagnetic fields (beyond the usual dielectric response and damage by ionizing gamma radiations). Extensive investigations were carried out with both prokaryotic and eukaryotic purified DNA samples utilizing some of the most sensitive and precision instrumentation and methods available, while scanning the whole spectral region from 1Hz to 100KHz (in the low frequencies) and all the way to the high-frequency region of 100MHz (including investigations on the effects of 100MHz high-frequency fields as well as 2.4GHz microwave fields on the DNA). We were unable to detect any electromagnetism of any kind intrinsic to the DNA or its coupling to external noise sources, whether concentrated or diluted in water, as compared to control samples ($H_2O$ or empty chambers), neither existence of any possible spontaneous or stimulated/induced electromagnetic fields or waves emanated from both the eukaryotic and prokaryotic genetic material. Based on our measurements, we conclude that either there is no intrinsic electromagnetic activity or fields present in the DNA material, in both concentrated or diluted form, or any such activity is extremely weak in its intensity and beyond the measurement limits of current scientific methods.

As per our current understanding of science and the contemporary physical models in both *classical* [1] and *quantum electrodynamics* [2], a material as the Deoxyribonucleic Nucleic Acid (DNA), similar to other polymers (or polymers with a biological function), in principle, should not have any significant intrinsic electromagnetic properties, intrinsic fields or interactions, except usual interactions with external electrical or electromagnetic fields with insignificant effects, which all other polymers and similar non-living materials exhibit. The long strings of these cylindrical polymers which carry the genetic code of life have long been debated for having possible permanent electric moments [3]. However, by virtue of the extremely minuscule electric field-induced dipole moments of the DNA, which enable them to align to external electric fields (as seen in aqueous solutions) [1], and an induced dichroism observed in DNA fragments in the presence of high electric fields [4], some kind of weak DNA electrical properties may exist, in theory.

As far as intrinsic electromagnetic properties and interactions in the molecule are concerned, it remains a remote possibility, especially in view of the warm temperatures and high damping within the cell nucleus and the cell cytoskeleton, precluding any quantum electrodynamic effects. If there were an intrinsic or induced activity in DNA, its strength might be much weaker than the limits of the current measurement instruments, especially to distinguish it from the ubiquitous noise present in the ambient environment of the DNA. Hydrogen bonds involved in DNA may act as dipoles and their movement could lead to spontaneous emission or absorption of electromagnetic radiation, however, the extremely weak magnitudes of ensuing fields and their damping by the complex, dissipative and warm biological environment would most probably lead to non-observable effects. In both cases, one cannot expect to observe any conspicuous electromagnetic effects within the DNA. A few theoretical models and arguments for electric properties and interactions of DNA have been reported in the literature [5], but theoretical models or pertinent experimental measurements of DNA electromagnetic properties are yet to be reported. The only large body of experimental data available related to DNA electromagnetic interactions is the radiation damage caused by gamma rays (the most energetic form of electromagnetic radiation), but the effect is more of an ionizing radiation-induced energy deposition and resulting damage rather than a pertinent electromagnetic interaction *per se*. So far, no appropriate scientific study has been carried out to investigate intrinsic bioelectromagnetism of the DNA, unlike cells of various kinds for which extensive scientific studies have been carried out.

It seemed logical and important to make careful measurements of any possible electromagnetic effects present in DNA by making use of some of the most sensitive instruments available in present times, which may shed light on this matter, as well as provide some introductory data to carry out realistic electromagnetic modeling of DNA, if at all an evidence of detectable electromagnetic activity could be recorded by current precision methods.

Before carrying out an investigation on the DNA, it would be imperative to contemplate its physical properties. As illustrated with the help of a cartoon, Figure 1, DNA is a polymer constituting double helix with a cross-section of approximately 2nm and a length ranging from few micrometers to centimeters in various organisms and is highly condensed *in vivo* [7]. Each helix is about 3.4nm in length and separation of about .34nm exists between each base pair. There is an inner elastic core which is surrounded by an outer hard core made of inelastic material. The circumference of the coil is 11nm and distance between each successive coil is about 2nm. Thus, in terms of electric dipole moments within these polymers and their possible dipole electromagnetic field emissions, the intensities of emanated fields should be extremely weak, much lower than what our instruments can measure. Besides, in view of the small dimensions ($l$) of these structures on the order of $10^{-10}$-$10^{-9}$m and much larger wavelengths ($\lambda$) of the expected fields on the order of $10^{1}$-$10^{3}$m, one would contend with poor radiation efficiency of any possible emanated fields which exists in the limit of $l \ll \lambda$ [17], making it quite difficult for these fields to be transmitted.

In order to carry out this study, we prepared and isolated seven samples of prokaryotic and eukaryotic DNA for our measurements. Samples 1 to 3, the eukaryotic DNA batch, was obtained from Mesenchymal Stem Cells (MSC) from murine bone marrow, which is one of the most viable models for MSC extraction [8, 9]. We prepared and isolated the samples following the protocols described elsewhere [10, 11], mainly using standard scientific-grade kits. Alternatively, for a more simpler extraction, the MSC DNA can also be obtained from dental pulp [12]. Samples 4 to 7 constituted the prokaryotic DNA batch, obtained from *E.coli* chromosomal DNA with the help of a standard commercial DNA isolation kit (Qiagen, www.qiagen.com). The protocols used were a modified form of the ones as found in the literature [13, 14]. The samples were suspended in a buffer and kept in DNA lock-seal vials (Fisher Thermoscientific) at 280K. For experiments, the samples were naturally thawed and immediately used in the experiments as soon as they reached the ambient room temperature. We used two kinds of sample preparations in our studies, at first, a DNA suspension in a buffer and second the same DNA suspension diluted in water.

Our experiments comprised two major strategies, first, to measure the sample's intrinsic electromagnetic response using an antenna in the proximity methodology, similar to RF radiometry, and second, using a direct-contact antenna method, similar to the electrometry methods. Suitable experiments for each method were carefully designed. The first suite of experiments (Figure 2) comprised a special micro-mesh copper coil antenna we devised for the purpose of measuring low to high frequency fields within the DNA. The antenna was developed using a mesh of 32SWG 99.99% pure copper wire covering the entire surface area of the vial, and the two ends were joined together to convey the signal. The signal obtained from the antenna was amplified with the help of a custom-made ultra-sensitive ultra-low-noise RF pico-ampere Trans-Impedance Amplifier (TIA), based on the femto-ampere input LMC6622 chip (National Semiconductor Corp.), operated with a regulated power supply based on batteries. The amplifier was calibrated with a 2182A Nano-volt Digital Multimeter with a precision source (Keithley

Corp.). The second suite of experiments (Figure 3), was quite similar to the first kind with the exception that the coil antenna was replaced by a copper electrode antenna which had a direct contact with genetic material kept in a petri dish. The output of the antenna was read out using the same procedure with our TIA as for the first suite of experiments.

All the experiments were performed at room temperature in a solid 99.99% pure copper grade I (grounded) Faraday cage (covered completely with an extra synthetic fiber electromagnetic shield canopy (Aaronia GmBH), however the amplifier and the antenna were pre-cooled with Liquid Nitrogen to temperatures around 100K to reduce the measurement noise before introducing the sample (which was kept at room temperature), in a way to increase the experiment sensitivity further. The Faraday cage was on an Anti-vibration table and all possible measures were taken to isolate the experiments from environmental noise. During the tests it was revealed that our instruments were extremely sensitive to even noise in the adjoining rooms and corridors, thus the actual data taking was carried out at night to reduce vibrations and other interference from human mechanical and electromagnetic activities around our laboratory in the day-time.

Before carrying out each experiment, proper calibration and test runs were performed, using data taking from empty chambers and chambers with pure water ($H_2O$). After the calibrations and control runs were performed, actual experiments were carried out under the identical conditions as control except with actual DNA samples. To supplement our spontaneous electromagnetic properties measurements we carried out induced signal measurements by irradiating the samples with 100MHz High-Frequency (HF) and 2.4GHz microwave fields. Data thus obtained was obtained, archived and analyzed with the help of a commercial high-precision 16-bit 2MHz BNC Data Acquisition System (MCCDAQ-1604HS2, Measurement Computing Corp.).

After the experiment and the measurement system were set up, a 50Hz frequency -20dBm intensity signal was injected into the cavity to calibrate the instrument. The response measured as a result of the signal is illustrated in Figure 4. A 50 Hz peak and its higher harmonics were observed all the way to 100KHz band. If there were any possible electromagnetism within the DNA, it would have been in some form similar to the response seen here.

A glimpse of our measurements made with two batches of the DNA are illustrated in the Figures 5A and B, which are the log-log plots made for convenience to accommodate the large span of frequencies. The response includes the concentrated DNA as well as diluted form in $H_2O$. A linear-linear raw results plot is illustrated in Figure 6 to depict what was actually measured in one of those samples. No response was seen from the DNA samples corresponding to either an intrinsic response or coupling to mains noise as seen in the 50Hz calibration signal.

Figure 7 illustrates a summary plot of all the seven samples. There is no response seen at all which could be beyond the noise with a good Signal-to-Noise Ratio (SNR), translating to around 2 or 3 at a minimum. We see no difference from the water sample, similar noise dominates the plots. Any pertinent intrinsic signal seen would have been of at least with an amplitude of around 0.001-0.003 (in arbitrary units).

What is observed in the recordings are a usual distribution of ubiquitous noise and signal power density as a function of frequency. The highest amplitude of the noise power is near the *dc* or 0Hz region, which corresponds to the low-frequency or *1/f* noise, which is associated with the measurement circuitry. Finally, what is seen at higher frequencies is the usual shot noise, as expected with electromagnetic measurements and thermal fluctuations. The peaks seen at 50/60Hz are the background interference from the mains noise, the peaks seen around 120Hz and to about few KHz are its higher harmonics. No discernible electromagnetic response is observed in any of the recordings, in short. The response is identical to a sample of ordinary buffer or water without the DNA in it.

Figure 8 depicts the response of the eukaryotic DNA sample to irradiation with strong (-50dBm) 2.4GHZ fields similar to "WiFi" devices. There is an absence of any possible extraordinary response seen with significant SNR or conspicuous intensity.

The observations reveal random noise which is dominated by mains and 1/f noise till about 500Hz (which are the primary peaks) followed by higher harmonics of the primary noise peaks, ie their multiples. There was no intrinsic bioelectromagnetism recorded from any of the DNA samples after extensive testing and multiple runs involving each batch. We neither saw a response intrinsic to DNA nor a coupling of DNA to external electromagnetic fields or noises where the sample could enhance the external stimuli. There was a flat response all the way from the low frequency spectral region to the high frequency. There was no significant event recorded. All events recorded which differed from the noise were owing to some kind of stimulus in the environment such as surrounding mechanical vibrations around our laboratory.

On the basis of our observations we surmise that the response seen appears to be the systemic noise from the experiments apparatus and in some cases a dielectric response of DNA, similar to other non-living biological material, resulting from the coupling of a sample to the environment. As expected of a polymer, DNA does not seem to exhibit any intrinsic bioelectromagnetism. Possibly, *in situ*, DNA could have more profound intrinsic activity once inside the functioning nucleus, owing to various conformations and dipole motions etc., albeit of very short-range nature and on small time scales owing to the highly damp and hot environment of the cell and nucleus, precluding its detection. However, once a DNA is purified and isolated from its natural environment, it does not seem to have a significant electromagnetic activity, except possibly acting as a complex antenna, which has been suggested elsewhere [15, 16]), whereby DNA, similar to other polymers, absorbs and re-transmits (in short reflects), any incoming electromagnetic waves like an antenna. Our observations may possibly support that reasoning.

Even if there were a bioelectromagnetism existent within purified DNA, its amplitude would be extremely faint and not possible to be measured using human methods, including some of the most sensitive and highest precision methods available today.

We suggest that mitochondria might possibly be a better candidate for probing intrinsic bioelectromagnetism within living material, as compared to a pure bio-polymer like DNA.

**Acknowledgments:**

We would like to acknowledge the support by the KACST (under grant number 234-35) and a laboratory development grant from the Jazan University Deanship of Scientific Research (DSR) which made this research possible.

---

Figures:

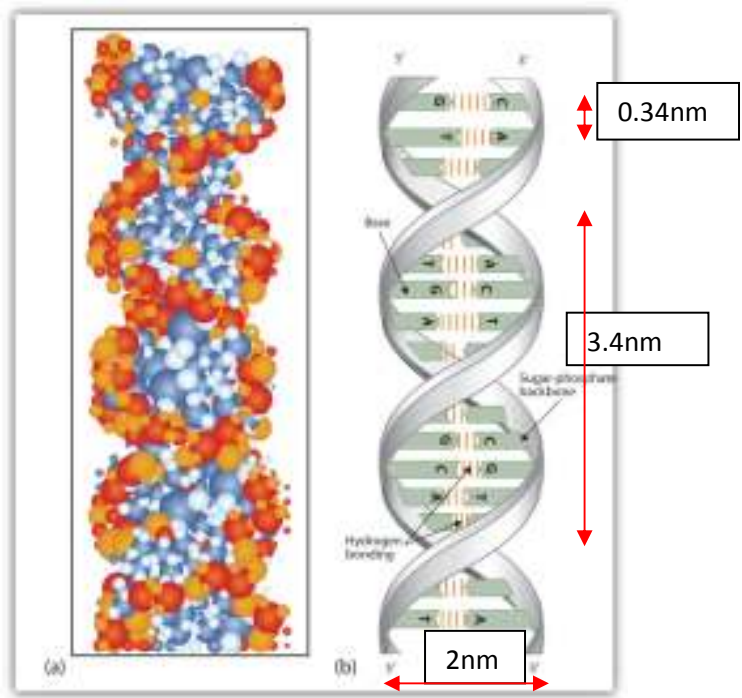

Figure 1: Dimensions of various segments of DNA super-imposed on a public-domain image of a DNA Double helix structure, in order to illustrate the scale of these structures and their effect on the underlying electromagnetic activity. (Image courtesy [6], with thanks.)

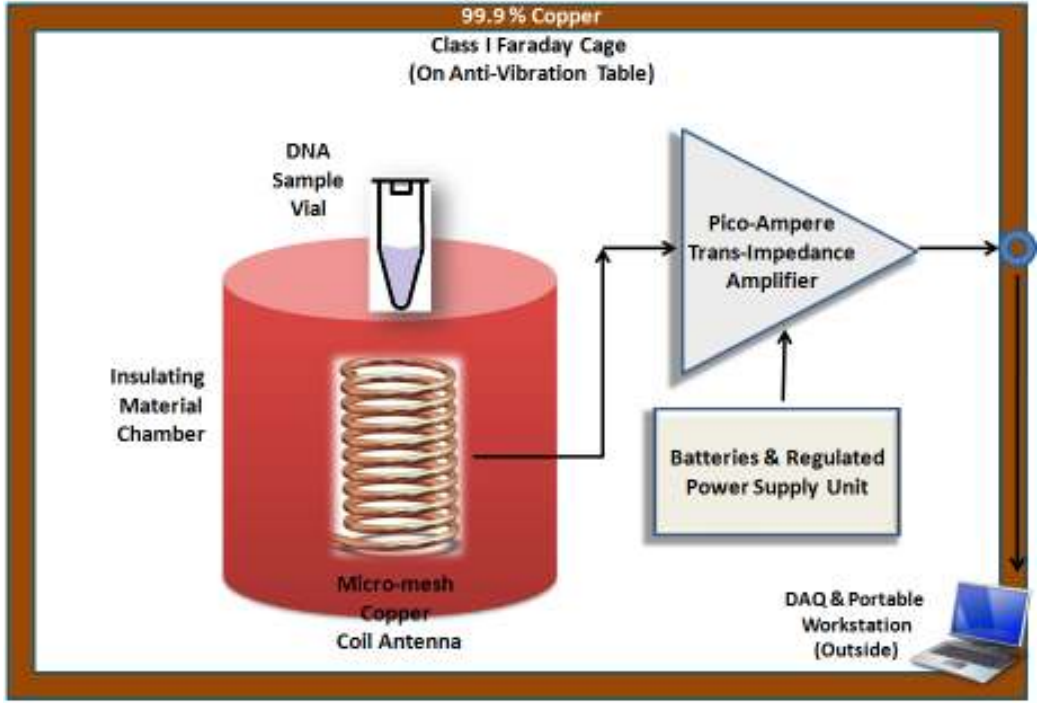

Figure 2: An overview of the technique devised and used in the first batch of experiments, the external coil antenna method.

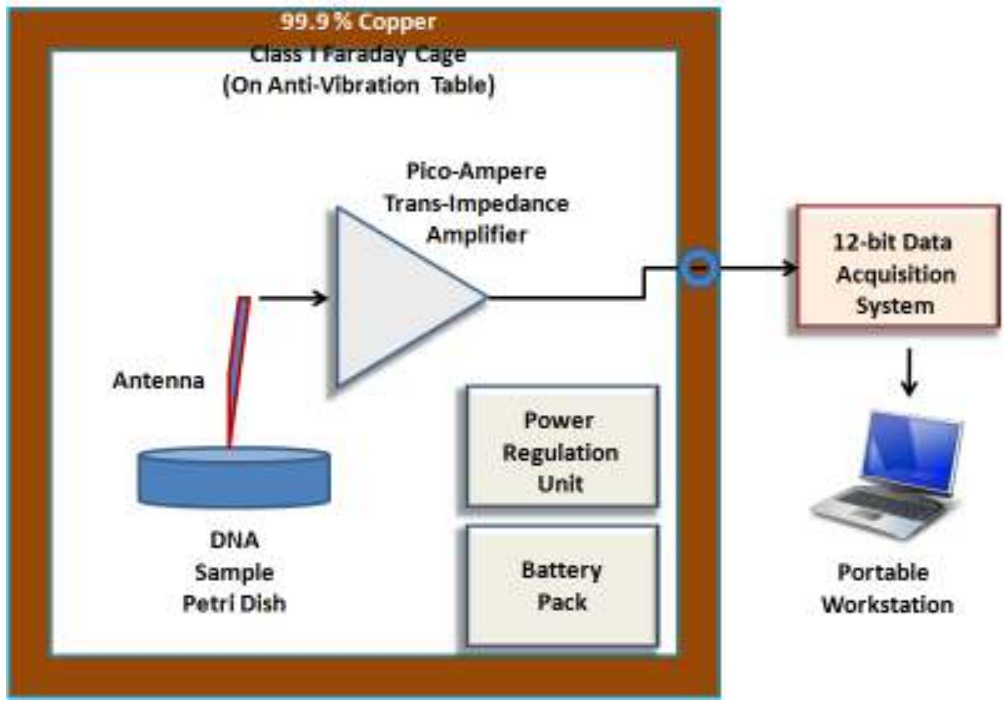

Figure 3: The methods used in the second batch of experiments, our so-called "Direct Contact" method, where an antenna sensed any possibly fields from the sample by means of a direct contact with it.

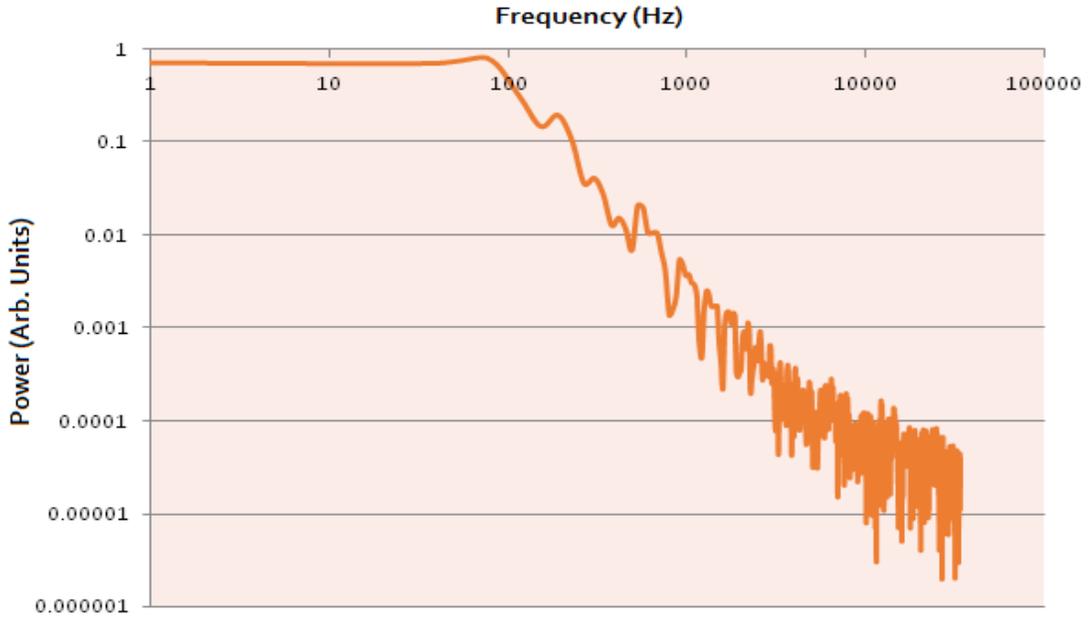

Figure 4: A calibration plot which illustrates the response of the instrument and measurement chamber with $H_2O$ to externally-generated 60Hz fields.

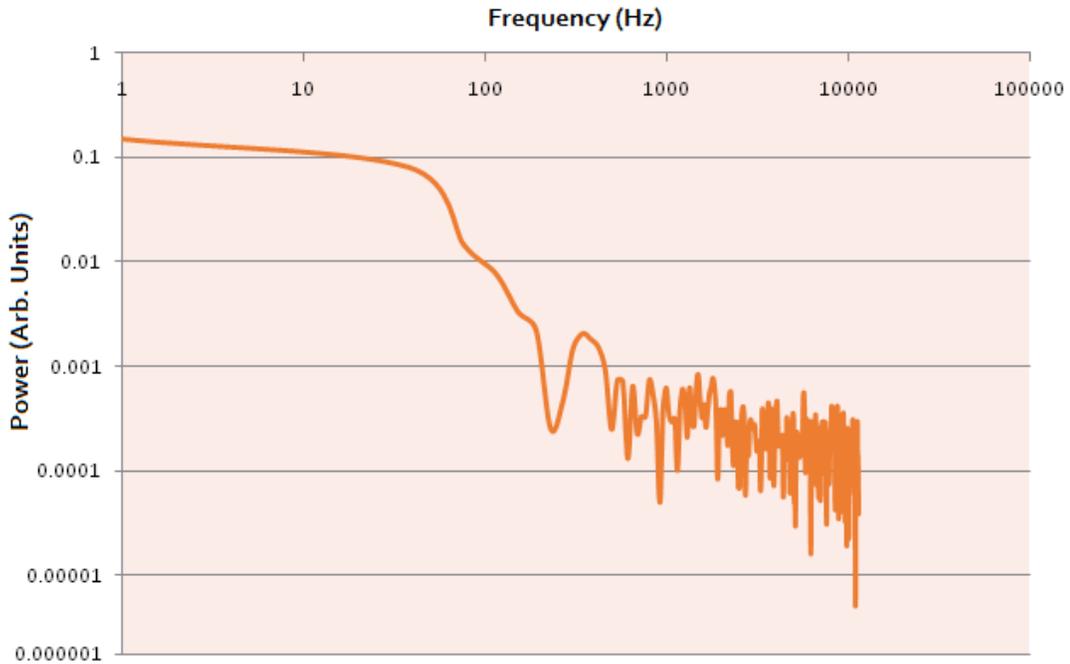

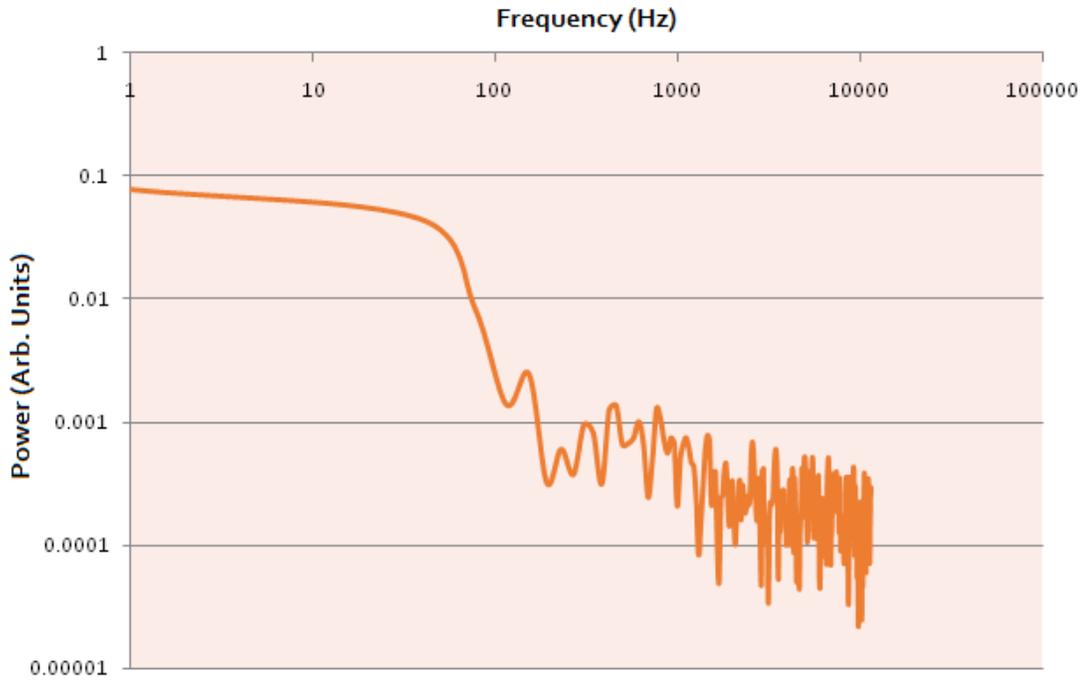

Figure 5a and b. The noise seen from two different batches of DNA. The spectral form of the noise seen is similar, there is only a 1/f and thermal noise seen, while no activity is seen at higher frequencies upto approximately 1-25KHz, the most anticipated region for any electromagnetic activity in the biological material.

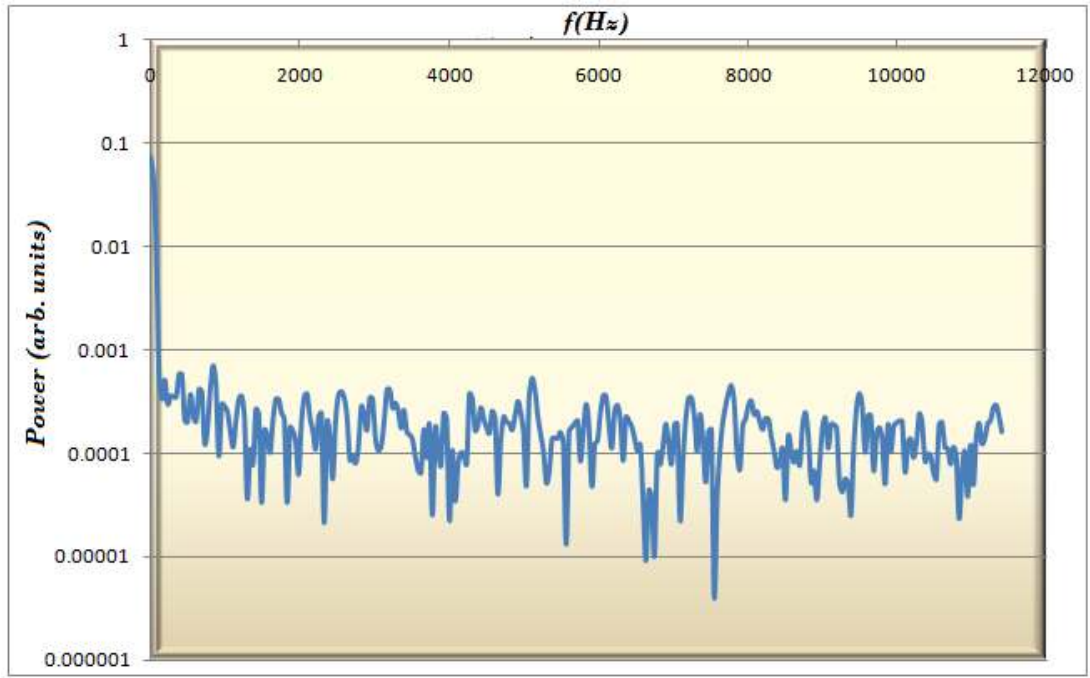

Figure 6: A linear plot of the raw signal measured from the DNA sample. No electromagnetic response intrinsic to the genetic material is seen except the usual thermal and 1/f noise fluctuations.

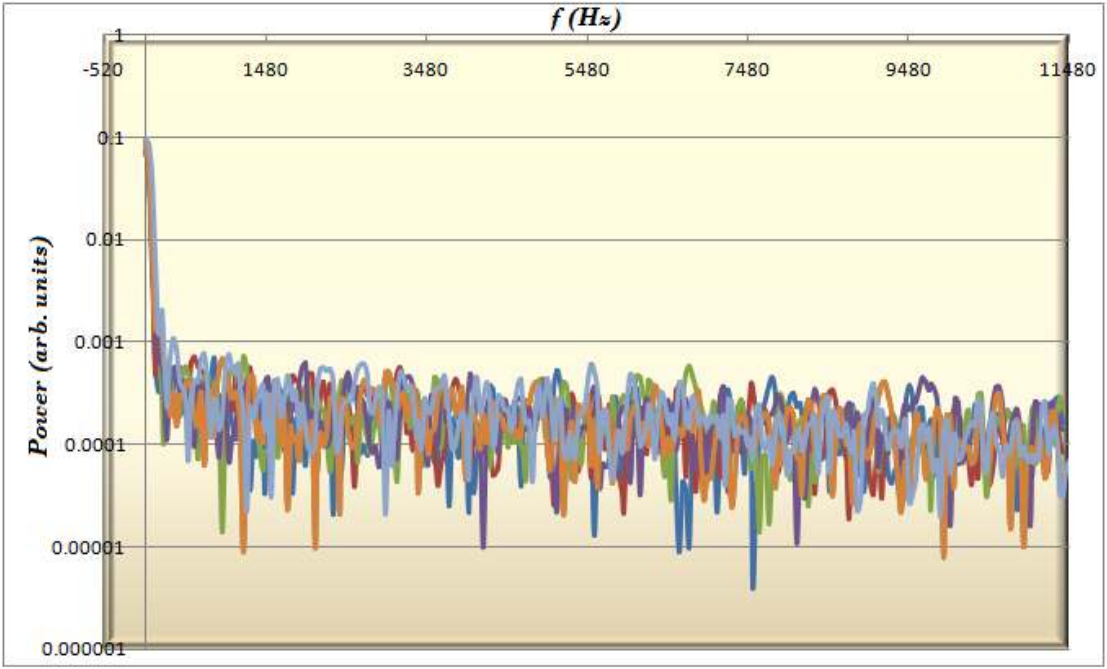

Figure 7: A summary plot of the data obtained from the seven samples used in our studies (three samples of eukaryotic DNA and four samples of prokaryotic DNA).

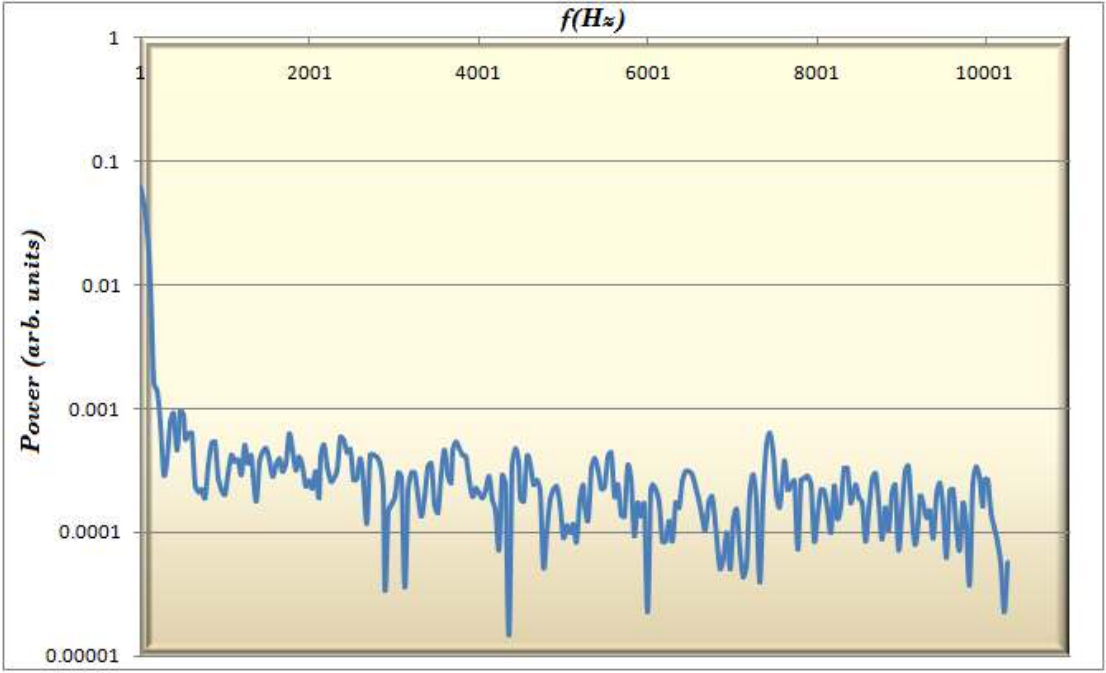

Figure 8: Response of DNA suspension to 2.4GHz microwave fields, depicting no response intrinsic to DNA as a result of irradiation.